\documentclass[conference]{IEEEtran}
\IEEEoverridecommandlockouts
\usepackage{cite}
\usepackage{amsmath,amssymb,amsfonts}
\usepackage{algorithmic}
\usepackage{graphicx}
\usepackage{booktabs}
\usepackage{textcomp}
\usepackage{xcolor}
\usepackage{array}
\usepackage{multirow}

\usepackage{hyperref} 
\hypersetup{
    linkbordercolor={0 0 1}
}
\newcolumntype{P}[1]{>{\centering\arraybackslash}p{#1}}

\newcommand{\rom}[1]{\uppercase\expandafter{\romannumeral #1\relax}}

\def\BibTeX{{\rm B\kern-.05em{\sc i\kern-.025em b}\kern-.08em
    T\kern-.1667em\lower.7ex\hbox{E}\kern-.125emX}}
\begin{document}

\title{Low-cost Active Dry-Contact Surface EMG Sensor for Bionic Arms\\
\thanks{*These authors contributed equally to the work. \newline \textcopyright \hspace{1pt} 2020 IEEE. Personal use of this material is permitted. Permission from IEEE must be obtained for all other uses, in any current or future media, including reprinting/republishing this material for advertising or promotional purposes, creating new collective works, for resale or redistribution to servers or lists, or reuse of any copyrighted component of this work in other works.}
}


\author{\IEEEauthorblockN{Asma M. Naim$^*$}
\IEEEauthorblockA{\textit{Dept. Electronic and Telecommunication} \\
\textit{Engineering, University of Moratuwa}\\
Sri Lanka \\
asmanaim@ieee.org}
\and
\IEEEauthorblockN{Kithmin Wickramasinghe$^*$}
\IEEEauthorblockA{\textit{Dept. Electronic and Telecommunication} \\
\textit{Engineering, University of Moratuwa}\\
Sri Lanka \\
kithminrw@ieee.org}
\and
\IEEEauthorblockN{Ashwin De Silva}
\IEEEauthorblockA{\textit{Dept. Electronic and Telecommunication} \\
\textit{Engineering, University of Moratuwa}\\
Sri Lanka \\
ashwind@ieee.org}
\and
\IEEEauthorblockN{Malsha V. Perera}
\IEEEauthorblockA{\textit{Dept. Electronic and Telecommunication} \\
\textit{Engineering, University of Moratuwa}\\
Sri Lanka \\
malshav@ieee.org}
\and
\IEEEauthorblockN{Thilina Dulantha Lalitharatne}
\IEEEauthorblockA{\textit{Department of Mechanical Engineering,} \\
\textit{University of Moratuwa}\\
Sri Lanka \\
thilinad@uom.lk}
\and
\IEEEauthorblockN{Simon L. Kappel}
\IEEEauthorblockA{\textit{Dept. Electronic and Telecommunication} \\
\textit{Engineering, University of Moratuwa}\\
Sri Lanka \\
simon@lkappel.dk}
}

\maketitle

\begin{abstract}
Surface electromyography (sEMG) is a popular bio-signal used for controlling prostheses and finger gesture recognition mechanisms. Myoelectric prostheses are costly, and most commercially available sEMG acquisition systems are not suitable for real-time gesture recognition. In this paper, a method of acquiring sEMG signals using novel low-cost, active, dry-contact, flexible sensors has been proposed. Since the active sEMG sensor was developed to be used along with a bionic arm, the sensor was tested for its ability to acquire sEMG signals that could be used for real-time classification of five selected gestures. In a study of 4 subjects, the average classification accuracy for real-time gesture classification using the active sEMG sensor system was 85\%. The common-mode rejection ratio of the sensor was measured to 59 dB, and thus the sensor's performance was not substantially limited by its active circuitry. The proposed sensors can be interfaced with a variety of amplifiers to perform fully wearable sEMG acquisition. This satisfies the need for a low-cost sEMG acquisition system for prostheses.
\end{abstract}

\begin{IEEEkeywords}
Surface electromyography, Active electrode, Flexible printed circuit, Bionic arm, Gesture classification 
\end{IEEEkeywords}

\section{Introduction}
Amputation is the removal of a limb by trauma, medical illness or surgery. There are approximately 10 million amputees in the world, of which nearly $30$\% are living with an upper extremity amputation \cite{c1}. Transradial amputations (forearm) account for $47$\% of all upper extremity amputations \cite{c2}. In recent years, diseases and accidents, both vehicular and work-related, have drastically increased the number of catastrophic injuries, resulting in limb losses. 

The rejection rate for prosthetics are high among upper limb amputees, as people sustaining upper limb amputations present complex rehabilitative needs. Proper rehabilitation and comfortable, affordable and functional prostheses are a huge benefit in the facilitation of functional restoration. Due to the high complexity of myoelectric transradial prosthetics, the commercially available prosthetics in this category are currently extremely costly (as high as \$ \!75,000) \cite{c3}. These prosthetics translate muscle activity into information which is used by motors to control the movements of the artificial limbs.

The muscle activity associated with finger movements is caused by variations in ionic currents of relevant muscle fibers, and can be measured as myoelectric potentials on the surface of the forearm. Surface electromyography (sEMG) is a method of acquiring myoelectric signals from the surface of the skin. sEMG is an important tool for Human-Computer interaction tasks, such as finger gesture recognition \cite{c4}. Typically, multiple sEMG sensors are placed on the forearm of a subject to enable the characterization of movements involving several muscles. Ideally, the sensors are placed on the skin surface directly above the muscle of interest, to obtain the highest quality sEMG signals \cite{c5}.  

sEMG signals are recorded using active or passive electrodes \cite{c6}. In active electrodes, the sEMG signals are amplified close to the source by the appropriate electronic circuitry located in the electrode assembly. In passive electrodes, no amplification is performed close to the electrode. Instead, the electrode material is connected directly to the sEMG amplifier, with a lead wire. 

Passive electrodes have been widely used in previous studies \cite{c7}. They are cheap, but generally more prone to noise interference, because the high impedance electrode signal is transmitted in lead wires, connecting the electrode to the amplifier. 

The interface between the skin and the electrode, can be either wet or dry. With wet electrodes, gel is applied between the skin and electrode to  improve the stability and reduce the impedance of the electrode to skin interface \cite{c8}. With dry-contact electrodes, no gel is applied between the electrode and skin. Thus, when using dry-contact electrodes, no skin preparation is needed \cite{c9}, and dry-contact electrodes are therefore very suitable for prolonged measurements of bio-electric signals. Here, an active electrode design is typically required to handle the higher impedance of the electrode to skin interface when compared to wet electrodes.

A major portion of the extensive research that has been done on sEMG acquisition electrodes focus on active electrodes, but rarely have researchers considered the mechanical design and fastening methods of the electrode for continuous use over long periods of time. Generally, the quality of the acquired bio-signals are higher when using active electrodes, as compared to passive electrodes [10, 11]. Ribeiro et al. introduced a dry-contact, active flexible electrode, for wearable bio-signal recordings, with a novel interface material that is highly bendable and comfortable for the wearer \cite{c12}. The electrode was designed to have the interface material deposited directly on the back of the flexible printed circuit (FPC) board. Tests conducted with this dry active flexible electrode, showed that it had better electrical characteristics than the traditional Ag/AgCl electrode, namely less power line interference and better response in the signal band.

Guerrero et al. studied a dry-contact electrode for acquiring multi-channel sEMG signals using three parallel gold-plated rod electrodes fixed onto a printed circuit board (PCB) \cite{c13}. The common-mode rejection ratio (CMRR) of each electrode was boosted by an independent driven right leg (DRL) circuit to obtain measurements of a higher quality. The dry, rigid electrode could accidentally detach from the skin more easily than a wet or flexible electrode. Thus, the failure of one electrode could compromise the entire set of measurements. 

Studies have also been performed on active sensors where surface mounted components were directly attached to a textile screen-printed circuit using polymer thick film techniques, for acquisition of ECG signals \cite{c14}. Merritt et al. connected passive electrodes to a buffer circuit screen-printed onto the fabric, to decrease the vulnerability of the signal to external interference. Although these electrodes were developed to adapt to the contours of the human body and acquire bio-signals of high quality; mechanical design considerations, like ease of attachment to the limb, were not considered.

This paper presents a novel low-cost, flexible, active, dry-contact sEMG sensor with a mechanical design that was optimized to reduce motion artifacts and enable easy attachment to the forearm. The sensor was developed to be a part of a wearable sEMG acquisition system, which enables multi-channel recording of low noise sEMG signals. The signal quality of the developed sensor was evaluated experimentally with the real-time finger gesture recognition algorithm described by De Silva et al. \cite{c15}. In addition, the CMRR of the sensor was characterized to ensure that there are no substantial performance limitation of the sensor due to its active circuitry.

\section{Methods}

This section is divided into two subsections. Section \rom{2}-A describes the design of a wearable sEMG sensor and section \rom{2}-B describes the experimental setup for finger gesture recognition and for sensor characterization.

\subsection{Overview of the Wearable Surface EMG Sensor}

For the design of the  dry-contact sEMG sensor, the following factors were considered, to obtain high quality wearable recordings \cite{c16}.

\begin{itemize}
\item Characteristics of the electrode material and the amplification circuitry.
\item Attachment of the sensor, to obtain a stable electrode to skin interface.
\item Optimal placement of the sensors with minimal distance to the active muscle areas of interest.
\end{itemize}

To obtain a focal pickup area, a bipolar configuration was chosen over a monopolar configuration. The sensor was designed according to the SENIAM (Surface EMG for Non-Invasive Assessment of Muscles) standards \cite{c17} with the following characteristics:

\begin{itemize}
\item Shape of an Electrode: Circular
\item Size of an Electrode (Diameter): $10$ \!mm
\item Inter-electrode distance: $20$ \!mm
\item Material: Stainless steel
\end{itemize}

For bio-potential sensors, a key parameter to obtain high quality recordings, is a stable and low impedance electrode-skin interface. Unfortunately, dry-contact electrodes are known to have very high electrode to skin impedances ($>$100 \!k$\Omega$, i.e. few \!M$\Omega$’s) \cite{c18}. However, with a stable skin contact, impedance variations can be reduced and impedance mismatches between the electrodes can thereby be minimized. Stainless steel has dominating polarizable characteristics, and is therefore appropriate for recording sEMG, where frequency content below $25$ \!Hz have little relevance \cite{c6}. Moreover, stainless steel is an inert and durable material and can therefore be reused numerous times, without significant degradation.

\begin{figure}[b!]
\centering
\begin{minipage}[b]{1.0\linewidth}
  \centerline{\includegraphics[width=\columnwidth]{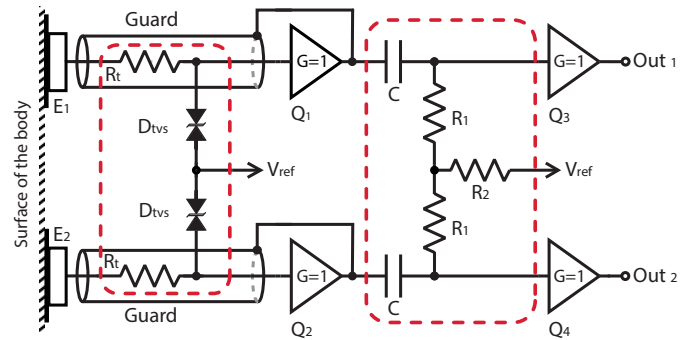}}
\end{minipage}
\vspace{-1.5cm}
\caption{Schematic of the active dry-contact sEMG sensor. The transient protection circuit and the differential high-pass filter are highlighted.}
\label{fig:schematic}
\end{figure}

The active circuitry of the sEMG sensor should have a high input impedance to accurately measure the sEMG signal. A high input impedance is necessary to obtain a high CMRR when there is a mismatch between the electrode-skin impedances of the attached electrodes \cite{c19}. The bias current and current noise flows through the source impedance (i.e. electrode-skin impedance), and must therefore be minimized to avoid a significant DC offset and noise contribution to the measured bio-potential.

Fig.~\ref{fig:schematic} shows the schematic for the active dry-contact sEMG sensor. The high impedance signals from the two electrodes, $E_1$ and $E_2$, are buffered by $Q_1$ and $Q_2$ and the buffered signals are used for active shielding of the electrodes. With active shielding, the low output impedance of the buffer drives the shield of the electrode. Thereby, only a negligible potential difference appears between the electrode signal and the shield, causing the displacement current to be close to zero. This design optimally protects the high impedance electrode signal from electrical interference \cite{c19}.

The buffered signals were high-pass filtered by a differential filter with a cut-off frequency at $15$ \!Hz, to stabilize the baseline of the signal. The output of the differential filter was buffered, to obtain a low impedance differential output, $Out_1$ and $Out_2$, of the electrode. The input of the buffers, $Q_1$ and $Q_2$, were protected by a transient voltage suppressor (TVS) diode at each input. 

The use of a unity gain buffer in the active circuitry means that the sensors are capable of being interfaced with commercially available, low-cost, open-source development platforms, such as the OpenBCI board \cite{c20}, for accurate and low-noise signal acquisition.

The buffers, $Q_1$ to $Q_4$, were implemented with the AD8244 (Analog devices, Massachusetts, USA). The AD8244 is a quad buffer with a high input impedance ($10$ \!T$\Omega$ $|$ $4$ \!pF), a low bias current ($2$ \!pA), and a very low current noise ($0.8$ \!fA$/\sqrt{\textit{{\normalfont Hz}}}$), which makes it suitable for buffering a high-impedance source. The TVS diodes were implemented with the TPD4E1B06 (Texas Instruments, Texas, USA) diode array. This diode array was specifically chosen due to its small line capacitance, $0.7$ \!pF, which is smaller than the $4$ \!pF input capacitance of the buffer, ensuring that the input impedance of the sensor is as high as possible.

\begin{figure}[b!]
\begin{minipage}[b]{1.0\linewidth}
  \centering
  \centerline{\includegraphics[width=\columnwidth]{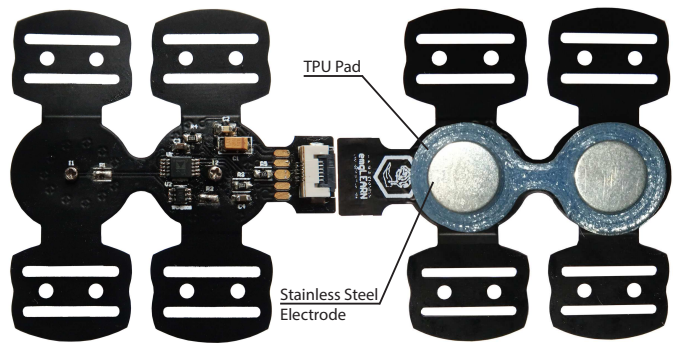}}
\end{minipage}
\vspace{-0.5cm}
\caption{(Left) Top side of the FPC with the components soldered. (Right) Bottom side of the FPC with stainless steel electrodes and a TPU pad around.}
\label{fig:flex}
\end{figure}

The design was fabricated and assembled on a double sided FPC, with a small form-factor. The cost of a single sensor was approximately 30 \!USD, including the stainless steel electrodes and the electronic components. Fig.~\ref{fig:flex} (Left) illustrates the top side of FPC with the components soldered, and Fig.~\ref{fig:flex} (Right) illustrates the bottom side of the FPC with the stainless steel electrodes attached and a thermoplastic polyurethane (TPU) pad around them.

\begin{figure}[h!]
\begin{minipage}[b]{1.0\linewidth}
  \centering
  \centerline{\includegraphics[width=\columnwidth]{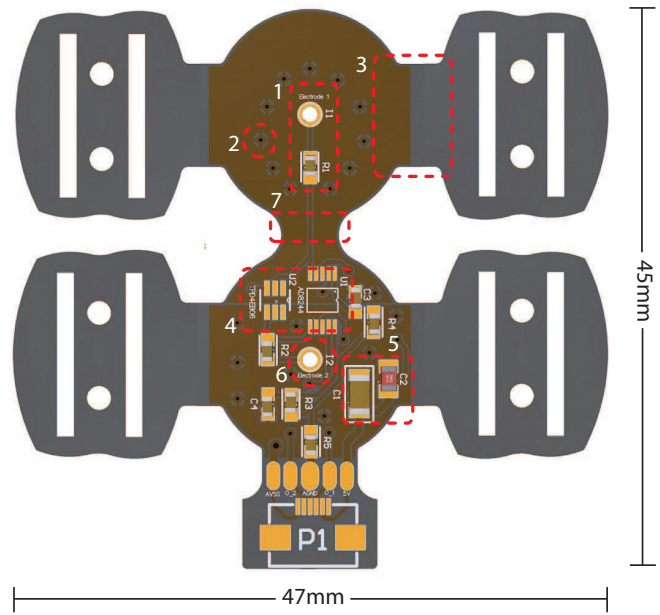}}
\end{minipage}
\vspace{-0.5cm}
\caption{The FPC layout with indications of the important design features. 
\newline1 -  High impedance input is actively shielded by copper filled planes. 
\newline2 -  Vias added to connect the shield planes on both sides of the FPC. 
\newline3 -  Flexible straps for bending support were excluded from the copper fill. 
\newline4 \& 5 -  Components were placed perpendicular to the direction of bending. 
\newline6 -  Shielded and plated through-hole to fix the screw to the electrode. 
\newline7 -  Narrow junction between the electrodes improve the flexibility towards rotation and bending.}
\label{fig:FPC}
\end{figure}

Fig.~\ref{fig:FPC} depicts the layout of the FPC sensor. The dimensions of the sensor are indicated on the figure, together with the most important design features, which are numbered and marked in dashed red. The sensor was designed such that it could be flexibly mounted onto the forearm using elastic straps, Fig.~\ref{fig:FPC}.3. All of the main electronic components were placed on the FPC surface covering the top side of one of the electrodes, Fig.~\ref{fig:FPC}.4, in order to utilize the semi-rigidity of this region of the FPC and to reduce the amount of traces crossing the narrow junction between the two electrodes. Fig.~\ref{fig:FPC}.4 and Fig.~\ref{fig:FPC}.5 show that all the components were placed laterally. This was done to reduce the risk of damaging the soldered components when bending the FPC around the forearm. In addition, the FPC has a narrow junction between the two electrodes, Fig.~\ref{fig:FPC}.7, to allow bending and twisting between the surfaces of the two stainless steel electrodes. The narrow junction of the FPC further allowed for slight rotations of the straps without affecting the area of contact between the skin and electrodes.

\vspace{-0.4cm}
\vspace{-0.3cm}

A mechanically stable contact between the electrode and skin is especially important when utilizing dry-contact electrodes, where a slight movement of the electrode can lead to significant variations in the impedance of the electrode to skin interface, resulting in motion related artifacts in the recorded data. To reduce these artifacts, a flexible pad, made of 3D printed TPU, was added around the electrodes, as shown in Fig.~\ref{fig:flex} . The flexible pad increased the skin contact creating a more stable electrode to skin interface.

\subsection{Experimental Setup}

Considering that the sensor was designed to be used along with a bionic arm, it was evaluated with a finger gesture classification experiment. In addition, the CMRR of the sensor was experimentally characterized.

\subsubsection{Real-Time Finger Gesture Classification}

The classification accuracy of the system based on the proposed sensor, was evaluated by placing sensors on the skin surface directly above the relevant muscles pertaining to the evaluated gestures and obtaining multi-channel sEMG recordings from each subject. The experiment was approved by the Ethics Review Committee at University of Moratuwa (Ethics Review Number: ERN/2019/007). 

\vspace{-0.2cm}
\begin{figure}[h!]
\begin{minipage}[b]{1.0\linewidth}
  \centering
  \centerline{\includegraphics[width=\columnwidth]{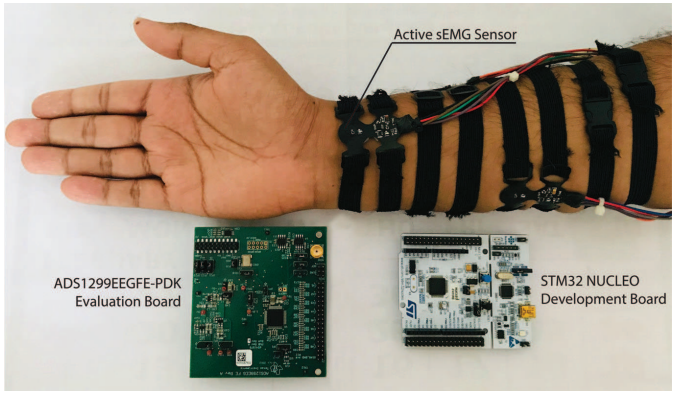}}
\end{minipage}
\vspace{-0.6cm}
\caption{The experimental setup for real-time finger gesture classification}
\label{fig:Class}
\end{figure}
\vspace{-0.1cm}
\begin{figure}[b!]
\vspace{-0.5cm}
\begin{minipage}[b]{1.0\linewidth}
  \centering
  \centerline{\includegraphics[width=\columnwidth]{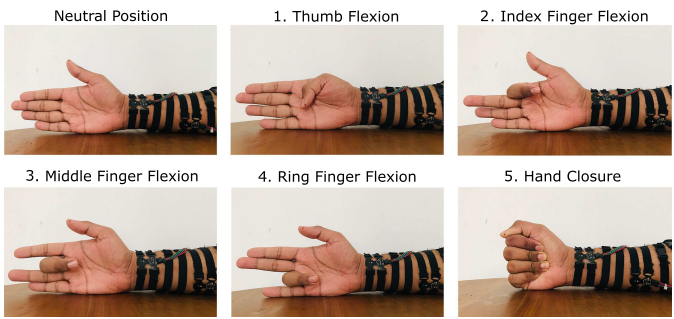}}
\end{minipage}
\vspace{-0.6cm}
\caption{Finger gestures. Top row:  In order from left to right; the hand in the neutral position, thumb flexion, index finger flexion. Bottom row: In order from left to right; middle finger flexion, ring finger flexion, hand closure.}
\label{fig:res}
\end{figure}

In order to perform real-time finger gesture classification, four sensors were connected to a bio-potential amplifier, consisting of an ADS1299EEGFE-PDK Evaluation (EVM) Board (Texas Instruments, Texas, USA), and a STM32 NUCLEO-F411RE Development Board (STmicroelectronics, Texas, USA) as shown in Fig.~\ref{fig:Class}. In addition, a Ag/AgCl wet electrode was attached near the elbow to act as a bias electrode, connected to the Vref, where Vref was the mid-value of the supply voltage. The sensors were mounted using elastic straps and buckles, ensuring a stable and tight electrode-skin contact and a comfortable pressure on the forearm. 

The five gestures illustrated in Fig.~\ref{fig:res} were included in the experiment, as most commonly used gestures are a combination of these gestures. Raw sEMG were recorded from four healthy subjects (2 males and 2 females, age: $24 \pm 2$), at a sampling frequency of $250$ \!Hz, using four sensors placed at optimal forearm positions  according to Crepin et al.\cite{c5}.

During the data collection, the subjects were asked to perform 20 repetitions of each gesture with their dominant hand. Each gesture was held for a period of 5 seconds, followed by a resting period of 5 seconds, during which the subjects were asked to keep their hand in a relaxed neutral position. The classification algorithm uses temporal muscle activation (TMA) maps, and was trained and tested on an individual subject basis. The data collection protocol and classification algorithm are detailed in \cite{c15}.

\subsubsection{Electrical Characterization of the Electrode}

\begin{figure}[h!]
\begin{minipage}[b]{1.0\linewidth}
  \centering
  \centerline{\includegraphics[width=7.1cm]{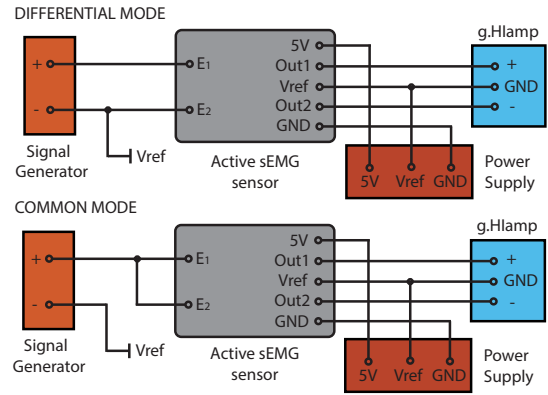}}
\end{minipage}
\vspace{-1.6cm}
\caption{The experimental setup for the electrical CMRR characterization }
\label{fig:CMDM}
\end{figure}

All sEMG signals are measured as differential signals between two electrodes. Therefore, common-mode signals, caused by e.g. electromagnetic interference, are considered as noise. Thus, a high CMRR of a bipolar sEMG sensor is crucial to obtain a good signal recording quality \cite{c24}. For the characterization of the sensor's CMRR, a signal generator was connected directly to the active sEMG sensor inputs, as shown in Fig.~\ref{fig:CMDM}. A g.HIamp amplifier (g.Tec Medical, Austria) was used to record the differential output from the developed sEMG sensor.

The sensor was placed in an electro-magnetic interference (EMI) shielded box, with wires drawn out for the power supply, inputs and outputs. The EMI shielded box was placed far from other electronic devices and the mains power supply to further reduce the noise interference. The sensor's electrical characteristics were determined via laboratory experiments. 

The power spectral density (PSD) at the sensor output was determined for both a differential-mode and a common-mode source, and the CMRR was the ratio between these PSDs as given in (1). The source was a signal generator, adjusted to a sine wave output of a fixed frequency, $\omega$. The signals were recorded for a fixed time period, and the recorded signals were used to determine the CMRR of the sensor, as given by (1).

\vspace{-0.1cm}
\begin{equation}
    CMRR (\omega) = 10\cdot
\log_{10} \left( \frac{PSD_d(\omega)}{PSD_c(\omega)} \right)
\end{equation}
\vspace{-0.2cm}

where, $CMRR(\omega)$ is the CMRR at the signal generator frequency $\omega$ and $PSD_d(\omega)$ and $PSD_c(\omega)$ are the PSD of sensor output at frequency $\omega$ for the differential-mode source and the common-mode source, respectively.

\section{Results and Discussion}

\subsection{Real-Time Finger Gesture Classification}
\begin{figure}[h!]
\begin{minipage}[b]{1.0\linewidth}
  \centering
  \centerline{\includegraphics[width=\columnwidth]{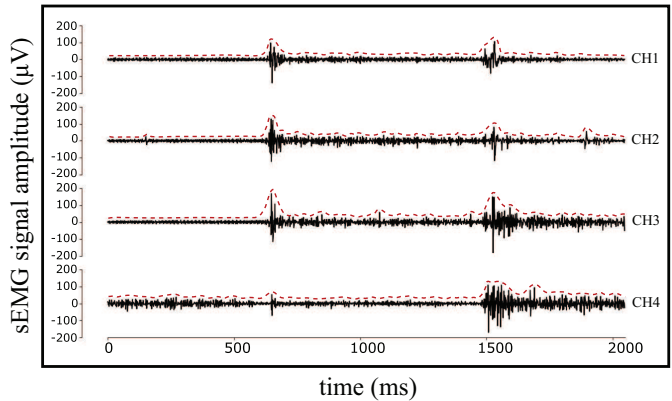}}
\end{minipage}
\vspace{-0.5cm}
\caption{Sketch of sEMG signal profiles (black) and signal envelopes (dashed red), for flexion and extension of the hand closure gesture, using four sensors.}
\label{fig:Signals}
\end{figure}

The results obtained from the experiment are summarized in Fig.~\ref{fig:Signals} and Table \rom{1}. Fig.~\ref{fig:Signals} shows the sEMG signals (black) obtained for flexion and extension of the hand closure gesture  using four sensors, along with the signal envelope used for real-time finger gesture classification (dashed red).

\vspace{-0.1cm}

Table \rom{1} reports the accuracies obtained for each finger gesture classification from the four test subjects. The sEMG signals obtained from four sensors could be used to classify the five finger gestures with an average accuracy of $85$\%.

\vspace{-0.2cm}
\begin{table}[h!]
\caption{Classification accuracy (\%)}
\vspace{-0.5cm}
\begin{center}
\renewcommand{\arraystretch}{1.3}
\begin{tabular}{|P{1.28cm}|P{0.32cm}|P{0.32cm}|P{0.32cm}|P{0.32cm}|P{0.4cm}|P{0.01cm}|P{0.93cm}|P{0.93cm}|cc}
\cline{2-6}\cline{8-9}
\multicolumn{1}{c|}{} & \multicolumn{5}{c|}{\textbf{Proposed sEMG Sensor}} & & \textbf{A. De} & \textbf{Crepin}\\\cline{1-6}
\multirow{2}{*}{\textbf{Finger}} & \multicolumn{4}{c|}{\textbf{Subject}} & \multirow{2}{*}{\textbf{Avg}} & & \textbf{Silva et} & \textbf{et al.}\\\cline{2-5}
 & \textbf{A}$^{\mathrm{a}}$ & \textbf{B} & \textbf{C} & \textbf{D} & & & \textbf{al. [15]} & \textbf{[5]} \\ \cline{1-6}\cline{8-9}

\multicolumn{1}{|l|}{$1$. Thumb$^{\mathrm{b}}$} & 80 & 90 & 80 & 90 & 85 & & -- & \textbf{91.91}\\
\multicolumn{1}{|l|}{$2$. Index$^{\mathrm{b}}$} & 90 & 100 & 80 & 90 & 90 & & -- & \textbf{95.38} \\
\multicolumn{1}{|l|}{$3$. Middle} & 80 & 90 & 90 & 80 & 85 & & \textbf{96.67} & 81.90 \\
\multicolumn{1}{|l|}{$4$. Ring} & 70 & 80 & 90 & 90 & 82.5 & & \textbf{94.58} & 93.50 \\
\multicolumn{1}{|l|}{$5$. Hand} & -- & 80 & 80 & 80 & 80 & & \textbf{93.75} & 77.87 \\ \cline{1-6}\cline{8-9}
\multicolumn{9}{l}{$^{\mathrm{a}}$Subject was a pilot study, and therefore gesture 5 was not measured.}\\
\multicolumn{9}{l}{$^{\mathrm{b}}$The classification of the finger gesture was not considered in \cite{c15} as}\\
\multicolumn{9}{l}{the muscles pertaining to the finger motion are not easily accessible by} \\
\multicolumn{9}{l}{the sEMG acquisition device used in the study.} \\
\end{tabular}
\end{center}
\end{table}
\vspace{-0.3cm}

Using the developed sEMG sensors, we were able to obtain classification accuracies above 80\%. The accuracies were generally lower than comparable results obtained by De Silva et al. \cite{c15} and Crepin et al. \cite{c5}. The lower accuracies could be caused by a lower signal quality, which might be related to the biopotential amplifier that was used to perform the study. The biopotential amplifier was based on an evaluation board, making it difficult to obtain optimal wiring and shielding of electrode cables and traces on the PCB. Future studies should test the developed electrode with a commercial grade EMG amplifier.

Another aspect that might be possible to improve, is the mounting of the sensor. The selected mounting method was a trade-off between signal quality, usability, and ergonomics. With additional user testing,  the sensor design can most likely be improved to obtain a better stability of the electrode to skin interface, and thereby improving the signal quality.

\subsection{Electrical Characterization of the Sensor} 
\begin{figure}[h!]
\begin{minipage}[b]{1.0\linewidth}
  \centering
  \centerline{\includegraphics[width=\columnwidth]{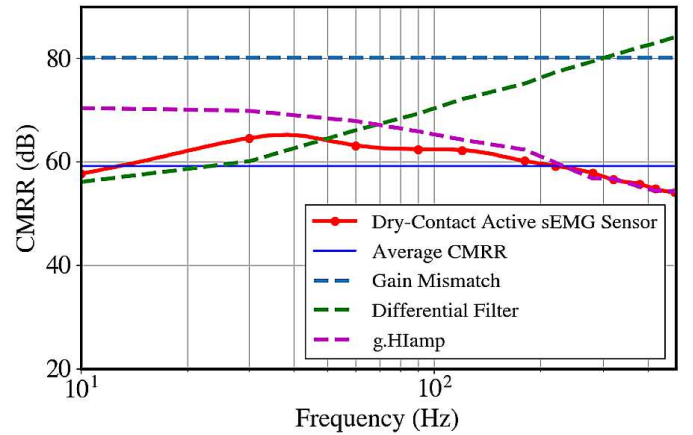}}
\end{minipage}
\vspace{-0.5cm}
\caption{The measured CMRR of the dry-contact sEMG sensor, compared to the CMRR of the buffer gain mismatch, differential filter, and bio-potential amplifier (g.HIamp).}
\label{fig:CMRR}
\end{figure}

The results of the CMRR characterization are summarized in Fig.~\ref{fig:CMRR}. The sensor was tested in the frequency range from $10$ \!Hz to $500$ \!Hz, corresponding to the sEMG signal bandwidth.

The CMRR of the sensor is limited by:
\begin{itemize}
\item The g.HIamp amplifier: The CMRR of the g.HIamp system was characterized by connecting the signal generator directly to the amplifier.
\item Differential high-pass filter: The bias resistor, $R_2$, of the differential high-pass filter had a value of 10 \!M$\Omega$. Increasing the value of $R_2$ improves the CMRR of the filter, and thus the value was a trade-off between the offset caused by the bias current and the degradation of the CMRR caused by the filter. For finger gesture classification, it was noted that a $10$ \!M$\Omega$ resistor was sufficient to obtain
good classification accuracies. 
\item Gain mismatch: Gain mismatches between buffers of the developed sensor deteriorate the CMRR, when the buffers are used to amplify a differential signal. 
\end{itemize}

Here, CMRR related to the gain mismatch and the differential filter were theoretical values. The gain mismatch was obtained from the datasheet of the buffer, whereas CMRR of the differential filter was based on equations derived by Casas et al. \cite{c27}. 

The CMRR obtained from the characterization experiment was sufficient to have an accurate classification of the five selected finger gestures. From $10$ \!Hz to approximately $50$ \!Hz, the CMRR was limited by the differential high-pass filter. From approximately $50$ \!Hz and above, there is a degradation in the measured CMRR due to the bio-potential amplifier, g.HIamp. 

\subsection{Mechanical Design of the Sensor}

Prior to the development of the FPC based sensor, an alternative approach based on a rigid PCB with TPU pads was also tested. Although the rigid PCB costs less than FPC, the inability of the sensor to bend along the contour of the forearm often resulted in one or both of the electrodes easily losing contact with the skin, which made the sEMG recordings unstable.

Moreover, slight movements of the arm resulted in motion artifacts in the observed signals. When one or both electrodes loses contact with the skin, there will be changes in the electrode-skin interface, which can result in very high source impedances and impedance mismatches. Thus, the FPC based sensor was chosen over the rigid PCB based sensor.

\section{Conclusion}

In this paper, a novel low-cost, active, dry-contact surface EMG sensor was introduced for use along with bionic arms. The sensor uses stainless steel electrodes to acquire signals from the surface of the skin. The proposed sensors can be interfaced with a variety of amplifiers, and this satisfies the need for a fully wearable, low-cost sEMG acquisition system, for prostheses development.

The developed sensor was evaluated for its ability to acquire high quality signals, related to five selected finger gestures, that could be classified in real-time. An average classification accuracy of $85$\% was obtained, using four sensors placed on the skin above the muscles corresponding to the selected gestures. 

The CMRR of the sensor was determined to ensure that the sensor's performance was not substantially limited by its active circuitry, but rather by other external factors. The CMRR for the developed sensor was measured to an average of $59$ dB.

The classification accuracy obtained using four sensors shows that the proposed active, dry-contact sEMG sensors can be used to obtain high quality, distinct signals for each finger gesture. The flexible sEMG sensors, along with optimal sensor placement, can be used to obtain high accuracy individual finger control for a bionic arm at a low cost.

Looking forward, it might be possible to further improve the sensor design to obtain a better stability of the electrode to skin interface, and thereby a better signal quality. The number of subjects should be increased to experimentally determine a more generalised method of mounting, in order to further improve the classification accuracies. Future work would also include the development of a fully wearable sEMG acquisition system, with the ability to accurately classify finger gestures and translate it to a bionic arm in real-time.

\section*{Acknowledgment}

Authors extend their gratitude to the Bionics Laboratory of Dept. of Mechanical Eng. at the University of Moratuwa.


\begin{thebibliography}{99}

\bibitem{c1} M. LeBlanc, ENGR 110, Class Lecture, Topic: Give Hope - Give a Hand, Rotary LN-4 Prosthetic Hand Project. University of Stanford, Stanford, California, November 9, 2008.
\bibitem{c2} D. W. Braza and J. N. Yacub Martin, Chapter 119: Upper Limb Amputations, Essentials of Physical Medicine and Rehabilitation (Fourth Edition), 2020, pp. 651-657.
\bibitem{c3} D. van der Riet, R. Stopforth, G. Bright, and O. Diegel, "An overview and comparison of upper limb prosthetics," IEEE 2013 Africon Conference, 2013, pp. 1–8.
\bibitem{c4} S. Benatti, B. Milosevicy, F. Casamassima, P. Schonle, P. Bunjakuz, S. Fatehz, Q. Huangz, and L. Benini, "EMG-based hand gesture recognition with flexible analog front end," Proc. IEEE 2014 Biomedical Circuits and System Conference, BioCAS 2014, pp. 57-60.
\bibitem{c5} R. Crepin, C. L. Fall, Q. Mascret, C. Gosselin, A. Campeau-Lecours, and B. Gosselin, "Real-time hand motion using sEMG patterns classification," Proc. IEEE Engineering in Medicine and Biology Society Conference, 2018.
\bibitem{c6} J. G. Webster, Medical instrumentation: application and design, Fourth edition, John Wiley \& Sons, Hoboken, NJ, 2010. 
\bibitem{c7} Md. Tariquzzaman, F. Khanam, Md. Sohag, and M. Ahmad, "Design and implementation of a low-cost multichannel rectified EMG acquisition system," 19th International Conference on Computer and Information Technology (ICCIT), December 2016, pp. 261-265.
\bibitem{c8} S. L. Kappel, and P. Kidmose, "Study of impedance spectra for dry and wet EarEEG electrodes," Proc. Annual International Conference of the IEEE Engineering in Medicine and Biology Society, EMBS, 2015, Vol. 2015, November, pp. 3161–3164. 
\bibitem{c9} P. Laferriere, E. D. Lemaire, and A. D. C. Chan, "Surface electromyographic signals using dry electrodes," IEEE Transactions on Instrumentation and Measurement, Vol. 60, No. 10, pp. 3259-3268, Oct. 2011.
\bibitem{c10} O. T. Inan and G. T. A. Kovacs, "An 11 $\mu$w, two-electrode transimpedance biosignal amplifier with active current feedback stabilization," IEEE Transactions on Biomedical Circuits and Systems, Vol. 4, No. 2, pp. 93-100, April 2010.
\bibitem{c11} M. J. Burke and D. T. Gleeson, "A micropower dry-electrode ECG preamplifier," IEEE Transactions on Biomedical Engineering, Vol. 47, No. 2, pp. 155-162, February 2000.
\bibitem{c12} D. M. D. Ribeiro, L. S. Fu, L. A. D. Carlos, and J. P. S. Cunha, "A novel dry active biosignal electrode based on a hybrid organic-inorganic interface material," IEEE Sensors Journal, Vol. 11, No. 10, pp. 2241-2245, Oct. 2011.
\bibitem{c13} F. N. Guerrero and E. Spinelli, "Surface EMG multichannel measurements using Aative, dry branched electrodes," IFMBE Proceedings, pp. 1–4, 2015.
\bibitem{c14} C. R. Merritt, H. Troy Nagle, and E. Grant, "Fabric-based active electrode design and fabrication for health monitoring clothing," IEEE Transactions on Information Technology in Biomedicine, Vol. 13, No. 2, pp. 274-280, March 2009. 
\bibitem{c15} A. De Silva, M. V. Perera, K. Wickramasinghe, A. M. Naim, T. Dulantha Lalitharatne, and S. L. Kappel, "Real-time hand gesture recognition Using temporal muscle activation maps of multi-channel sEMG signals," 2020 IEEE International Conference on Acoustics, Speech and Signal Processing (ICASSP), Barcelona, Spain, 2020, pp. 1299-1303.
\bibitem{c16} B. Gerdle, S. Karlsson, S. Day, and M. Djupsjöbacka, Acquisition, Processing and Analysis of the Surface Electromyogram. Modern Techniques in Neuroscience. Chapter 26: 705-755. Eds: U. Windhorst and H. Johansson, Springer Verlag, Berlin, 1999. 
\bibitem{c17} H. J. Hermens, B. Freriks, C. Disselhorst-Klug, and G. Rau, "Development of recommendations for SEMG sensors and sensor placement procedures," Journal of electromyography and kinesiology: official journal of the International Society of Electrophysiological Kinesiology, 2000. ISSN: 1050-6411, Vol. 10, Issue. 5, pp. 361-74.
\bibitem{c18} S. L. Kappel, M. L. Rank, H. O. Toft, M. Andersen, and P. Kidmose, "Dry-contact electrode Ear-EEG," IEEE Transactions on Biomedical Engineering, 66(1), 150–158, 2018.
\bibitem{c19} E. Spinelli and M. Haberman, "Insulating electrodes: A review on biopotential front ends for dielectric skin-electrode interfaces," Physiological measurement. 31. S183-98, September 2010.
\bibitem{c20} OpenBCI, 2019. http://www.openbci.com 
\bibitem{c24} T. Uktveris and V Jusas, "Development of a modular board for EEG signal acquisition," Sensors 2018, No. 7: 2140.
\bibitem{c27} O. Casas and R. Pallas-Areny, "Basics of analog differential filters," IEEE Transactions on Instrumentation and Measurement, Vol. 45, No. 1, pp. 275-279, February 1996.

\end{thebibliography}
\end{document}